\documentclass{article}
\usepackage{frascatiphys}
\usepackage{graphicx}
\usepackage{lineno}
\begin{document}

\title{ 
 \uppercase{Prospects for Photon-Photon Measurements with Tagged Protons in ATLAS}}
\author{
Maciej Trzebi\'nski\\
{\em The Henryk Niewodnicza\'nski Institute of Nuclear Physics Polish Academy of Sciences}\\
{\em Radzikowskiego 152, 31-342 Cracow, Poland} \\
on behalf of the ATLAS Collaboration*
}
\maketitle
\baselineskip=11.6pt
\begin{abstract}
In 2017, ATLAS has been equipped with a new, dedicated detector system allowing measurements of forward protons scattered at small angles in diffractive and electromagnetic processes. These ATLAS Forward Proton detectors (AFP) can operate during the standard high-luminosity LHC runs and collect large amounts of integrated luminosity. This gives a possibility to study rare interactions, in particular, the two-photon processes. The physics programme includes measurements of photon-photon interactions present in the Standard Model, as well as using searches for new physics. In this paper, the AFP detectors and physics goals are briefly presented.
\end{abstract}
\baselineskip=14pt
%


\newcommand\blfootnote[1]{%
  \begingroup
  \renewcommand\thefootnote{}\footnote{#1}%
  \addtocounter{footnote}{-1}%
  \endgroup
}

\section{Introduction}
\blfootnote{* Copyright 2019 CERN for the benefit of the ATLAS Collaboration. CC-BY-4.0 license.}
In the majority of events of photon-photon and photon-proton scatterings at the LHC one or both outgoing protons stay intact. Since photon is a colourless object, such an exchange results in a presence of the rapidity gap between the centrally produced system and scattered protons. Thus, such events are of diffractive nature.

Diffractive processes are an important part of the physics programme at hadron colliders. This is also true for ATLAS\cite{ATLAS}, where a large community works on both phenomenological and experimental aspects of diffraction. In such events, a rapidity gap\footnote{Rapidity gap is a space in rapidity devoid of particles.} between the centrally produced system and scattered protons is present. Due to the exchange of a colourless object, a photon (in case of electromagnetic interaction) or Pomeron (strong force), one or both outgoing protons may stay intact.

\section{Detection Techniques}
The diffractive production may be recognized by the search for a rapidity gap in the forward direction or by the measurement of scattered protons. The first method is historically a standard one for the diffractive pattern recognition. It uses the usual detector infrastructure as trackers and forward calorimeters. Unfortunately, the rapidity gap may be destroyed by \textit{e.g.} particles coming from the pile-up -- parallel, independent collisions happening in the same bunch crossing. In addition, the gap may be outside the acceptance of a detector. In the second method, protons are directly measured. This solves the problems of gap recognition in the very forward region and a presence of pile-up. However, since protons are scattered at small angles (few hundreds microradians), additional devices called forward detectors are needed to be installed.

\subsection{ATLAS Forward Proton Detectors}
ATLAS is equipped with two sets of forward proton detectors: ALFA \cite{ALFA1, ALFA2} and AFP \cite{AFP}. ALFA (Absolute Luminosity For ATLAS) detectors are designed to measure the properties of the elastic cross-section, soft diffraction and low-mass exclusive production. These topics are not in the scope of this report -- readers interested in details should see \textit{e.g.} Ref. \cite{ATLAS_elastic_7TeV, ATLAS_elastic_8TeV} (properties of the elastic scattering measured by ATLAS at $\sqrt{s} = 7$ and $8$ TeV), Ref. \cite{LHC_forward_physics} (general overview; soft diffraction), Ref. \cite{exc_pions} (exclusive pion production) or Ref. \cite{brem} (diffractive bremsstrahlung).

\begin{figure}[htb]
    \begin{center}
        {\includegraphics[width=0.9\linewidth]{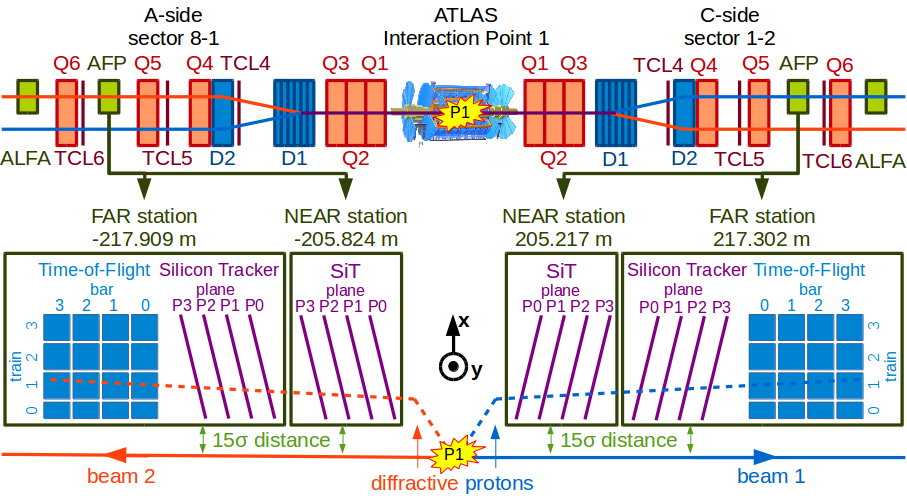}}
        \caption{\it Scheme of the ATLAS Forward Proton detectors.}
\label{AFP}
    \end{center}
\end{figure}

AFP (ATLAS Forward Proton) consists of four detector stations placed symmetrically with respect to the ATLAS Interaction Point at 205 m and 217 m. In each AFP station there is a Roman Pot device allowing the units to move horizontally. Detectors located on the ATLAS C side are inserted into the beam 1 whereas the ones on the A side into the beam 2. The scheme of AFP detectors is shown in Figure \ref{AFP}.

Each AFP station consists of four Silicon Trackers (SiT), which provide precise position measurements. The purpose of the AFP tracking system is to measure points along the trajectory of protons that were deflected during a proton-proton interaction. The readout chip was chosen to be FE-I4, which was originally designed for the IBL project \cite{IBL}. There are four such chips installed in each station. The active area covered by the tracking detector is approximately $16 \times 20$ mm$^2$ with a pixel size of $50 \times 250$ $\mu$m$^2$. Detectors are tilted by 14 degrees. The resolution of a single plane was measured to be about 6~$\mu$m in $x$ and about 30 $\mu$m in $y$ \cite{AFP_TB}. By having two detectors on each side of the Interaction Point (IP) one can measure not only the position of the proton with respect to the beam, but also its elevation angle. These are connected to the proton kinematics at the interaction point -- \textit{i.e.} by measuring the proton properties in the AFP one can unfold its initial four-momentum \cite{unfolding}.

Far stations host also the Time of Flight (ToF) detectors. The timing detectors measure the time of arrival of each proton, providing a trigger signal and allowing background reduction through the difference in proton time-of-flight measured on each side of the interaction point. The vertex calculated from ToF difference on both AFP sides can be compared to the primary interaction vertex. The resolution is expected to be between 20 and 30 ps \cite{AFP_TB}. 

\subsection{LHC Optics}
\begin{figure}[htb]
    \begin{center}
        {\includegraphics[width=0.9\linewidth]{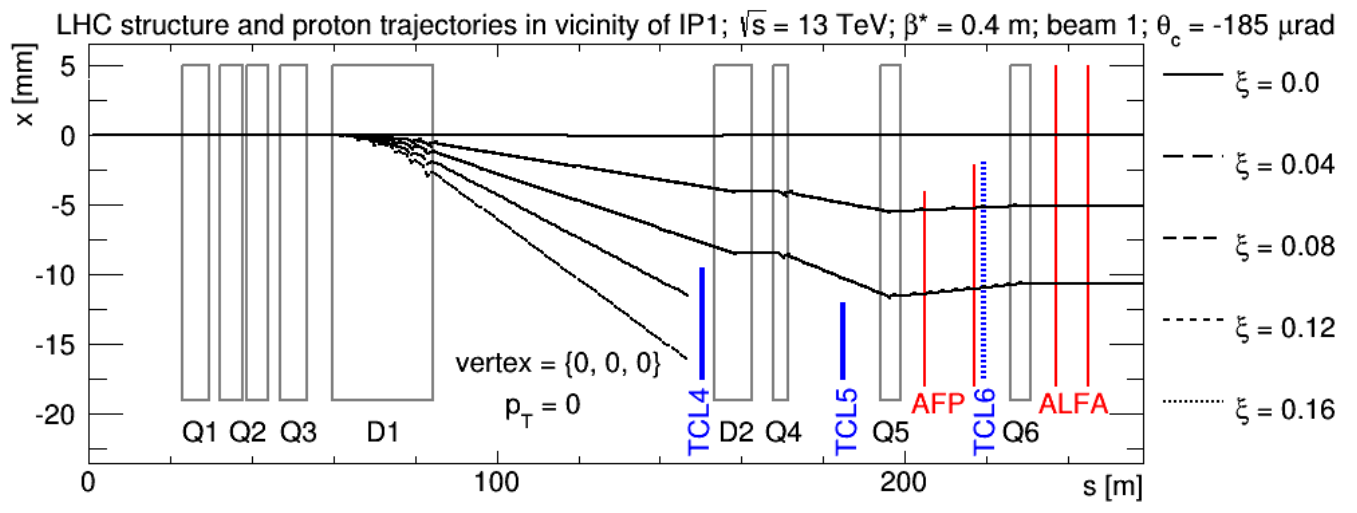}}
        \caption{\it Energy dependence of the proton trajectory in $(x, s)$ plane for $\sqrt{s} = 13$ TeV, $\beta^* = 0.4$ m LHC optics in vicinity of ATLAS Interaction Point (IP1). Protons were generated at $(0, 0, 0)$ with transverse momentum $p_\mathrm{T} = 0$. The crossing angle in horizontal plane was set to 185 $\mu$rad. For details see Ref. \cite{LHC_optics}.}
\label{optics}
    \end{center}
\end{figure}

Between the AFP stations and ATLAS interaction point several LHC elements are placed. They have a significant impact on the proton trajectory and will influence its position in the AFP stations. These elements are:
\begin{itemize}
  \item two dipole magnets (D1-D2) used for beam separation (bending),
  \item five quadrupole magnets (Q1-Q5) used for beam focusing,
  \item two collimators (TCL4, TCL5) used for magnet protection.
\end{itemize}
Settings of the LHC magnets are called optics and come from the requirements of the experiments in terms of luminosity and of the LHC machine protection. These settings may differ between the LHC fills. Due to the optics settings, the forward proton trajectories between the IP and AFP detectors are not straight
lines. A typical situation ($\beta^* = 0.4$ m optics) for the high-luminosity ATLAS data taking is shown in Figure \ref{optics}. Black rectangles represent dipole and quadrupole magnets, blue lines -- collimators and red lines -- AFP and ALFA stations. Assuming the proton transverse momentum equals zero, protons are bent accordingly to the energy lost during the collision: $\xi = \frac{E_{beam} - E_{proton}}{E_{beam}}$. As one can see, protons with very small energy loss are too close to the beam to be detected. With increasing energy loss, trajectories are further away from the beam and can be detected by AFP. However, if the energy loss is too big, forward protons will be filtered by collimators and will not reach the AFP station. The acceptance for a typical low-$\beta^*$ optics covers $0.025 < \xi < 0.1$ \cite{LHC_optics}, which corresponds to the proton energy loss of $160 < E_{proton} < 650$ GeV for $\sqrt{s} = 13$ TeV.

\section{Photon Physics with Forward Proton Measurement}
Photoproduction physics has so far been studied mainly in the electron accelerators. However, a high-energy bremsstrahlung from the proton at the LHC is a plentiful source of photons. Photoproduction processes can be studied using proton tagging.

\subsection{Two Photon Processes: $\gamma\gamma \to \mu\mu$}
\begin{figure}[htb]
    \begin{center}
        {\includegraphics[width=0.65\linewidth]{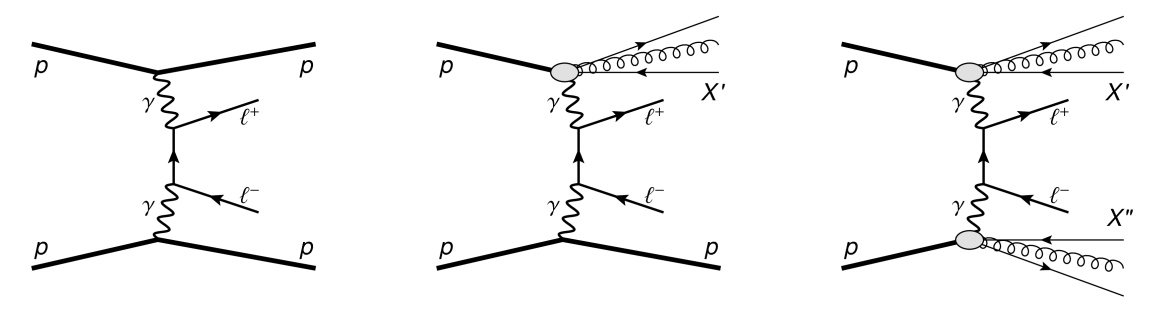}}
        \caption{\it Feynman diagrams for the exclusive di-muon photo-production.}
\label{diagrams_dimuons}
    \end{center}
\end{figure}

A di-muon system can be produced in the exclusive mode: $p + p \to p \gamma^* \gamma^* p \to p \mu^+ \mu^-p$, see Fig. \ref{diagrams_dimuons}. Such measurement was done by ATLAS without AFP \cite{ATLAS_muons}. The used data sample consisted of exclusive events and large irreducible background which was mainly coming from dissociated\footnote{Events in which one or both protons was destroyed due to interactions with other particles from the system. Such phenomena are described by a gap survival probability.} events. Information about the presence of scattered protons should allow a significant reduction of this background.

Due to the acceptance of AFP, double-tagged events have too small cross-section to be observed. However, a semi-exclusive (single tag) measurement should be possible assuming 40 fb$^{-1}$ of collected data, minimal muon transverse momentum of 10 GeV and AFP positioned at about 2 mm from the beam. As was discussed in Ref. \cite{AFP}, such measurement can be used for the AFP detector alignment and optics calibration.

\subsection{Anomalous Gauge Couplings}
\begin{figure}[htb]
    \begin{center}
        {\includegraphics[width=0.75\linewidth]{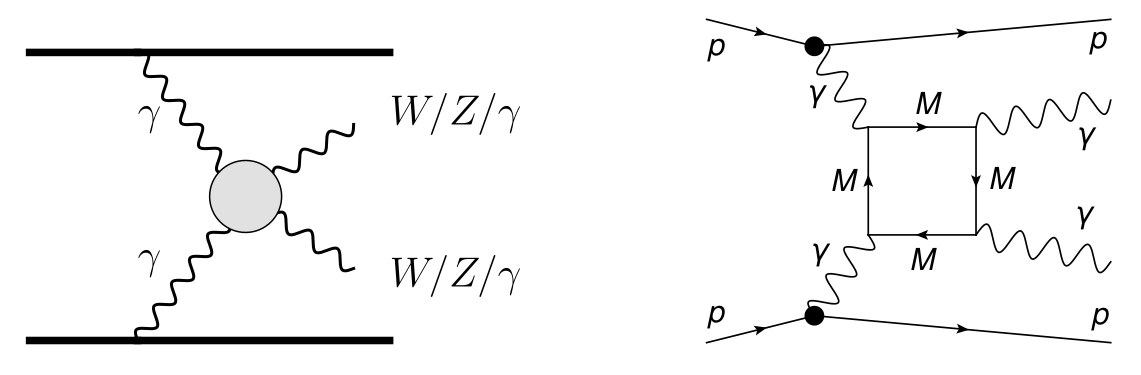}}
        \caption{\it Diagrams of anomalous gauge coupling (left) and magnetic monopole (right) production.}
\label{diagrams_AGC_MM}
    \end{center}
\end{figure}

Measurement of $W$ and $Z$ boson pair production via the exchange of two photons (see left panel of Fig. \ref{diagrams_AGC_MM}) allows to perform a stringent test of the electroweak symmetry breaking \cite{AGC1}. Standard Model predicts the existence of $\gamma \gamma W W$ quartic couplings while there is no $\gamma \gamma Z Z$ coupling. As was shown in Refs \cite{AGC2} and \cite{AGC3}, collecting 30 -- 300 fb$^{-1}$ of data with the ATLAS detector and using protons measured in AFP should result in a gain in the sensitivity of about two orders of magnitude over a standard ATLAS analysis.

\subsection{New Physics Searches}
Proton tagging may also serve as a powerful technique for the new physics searches as the backgrounds can be significantly reduced by the kinematic constraints coming from the AFP proton measurements. The general idea of background reduction was presented in Ref. \cite{excJJ1, excJJ2, excJJ3} on a basis of the exclusive
jet measurement.

Proton tagging technique might be also used for the invisible object searches. As an example, the case of magnetic monopoles produced by the photon exchange can be considered. From a diagram (see right panel of Fig. \ref{diagrams_AGC_MM}) one can conclude that, even if the centrally produced system escapes detection (or is not measurable) in ATLAS, one can measure scattered protons in AFP. In general, any production of new objects (with mass up to 2 TeV) via photon or gluon exchanges should be possible to be observed.

\section{Summary}

Since 2017 ATLAS is equipped with a full set of the AFP detectors, which collected data with a proton tag on both sides during the special and standard LHC runs. Even more data is planned to be collected during the LHC Run 3. Besides QCD measurements (rapidity gap survival, Pomeron structure, \textit{etc.}), photon-induced processes can be measured. These include single-tagged exclusive muons ($pp \to p \mu^+ \mu^-p$) and anomalous gauge couplings ($W$, $Z$ and photon pairs). For the latter processes the use of the AFP detectors provides a significant gain in sensitivity as compared to the measurement based on the data from the central ATLAS detector. On top of that one can try to search for any production of new objects produced via photon or gluon exchanges (magnetic monopoles, invisible particles, ...). In such searches, forward proton measurements can be used for a background reduction.

\end{document}